\documentclass[aps,prd,twocolumn,groupedaddress,showpacs,showkeys]{revtex4-1}
\usepackage{graphicx}
\usepackage{dcolumn}
\usepackage{bm}
\usepackage{amssymb,amsmath,latexsym,amsfonts}
\begin{document}


\title{Speed of sound of a Bose--Einstein condensate with dipole--dipole interactions}

\author{B. Gonz\'alez-Fern\'andez}
 \email{belinkag@nucleares.unam.mx}\affiliation{Instituto de Ciencias Nucleares, Universidad Nacional Aut\'onoma de M\'exico,\\ A. Postal
70-543, C.P. 04510, M\'exico, D.F., M\'exico.}

\author{A. Camacho}
\email{acq@xanum.uam.mx} \affiliation{Departamento de F\'{\i}sica,
 Universidad Aut\'onoma Metropolitana--Iztapalapa\\
 Apartado Postal 55--534, C.P. 09340, M\'exico, D.F., M\'exico.}


\date{\today}

\begin{abstract}
In the present work the case of a chromium Bose--Einstein condensate
is considered. The model includes not only the presence of the
so--called contact interaction but also a long range and anisotropic
dipole--dipole interaction has been included. Some thermodynamical
properties are analyzed. For instance, the size of the condensate,
chemical potential, speed of sound, number of particles, etc., are
deduced. It will be shown that this dipole--dipole interaction
implies the emergence of anisotropy, for example, in the speed of
sound. The possible use of this anisotropy as a tool for the analyze
of dissipative mechanisms, for instance, Landau's criterion for
superfluidity, will be also discussed.
\end{abstract}

\pacs{67.85De, 67.85Jk}
\maketitle

\section{Introduction}
Many of the properties of ultra--cold quantum gases are determined
by the features of the interactions among the atoms of the
corresponding system. For instance, symmetry characteristics,
intensity, or range of the interaction play a relevant role in the
observed phenomena \cite{Pitaevski1, Griffin1}. The changes that may
appear, in this context, can be surprising. Indeed, for attractive
short range interatomic interactions a Bose--Einstein condensate
(BEC) is unstable with regard to local collapses \cite{Dalfovo1}.
Nevertheless, the presence of a trapping potential allows, under
certain conditions, the existence of metastable states
\cite{Bradley1}. In other words, the system goes from unstable to
metastable with the introduction of a trapping potential. This
isotropic short range interaction, based upon an effective contact
interaction, can be tuned resorting to Feshbach resonances and many
interesting quantum features of these systems have been analyzed
with the use of this tuning possibility \cite{Griesmaier1}.

The realization of a BEC resorting to 52 Cr atoms \cite{Griesmaier1}
opens up an interesting window in the context of ultra--cold quantum
gases. Indeed, the large magnetic moment that some atomic species
possess, among them 52 Cr, offers the possibility of obtaining a
dipolar degenerate quantum gas \cite{Stuhler1}. One of the
advantages of 52 Cr comprises the fact that they have a dipole
moment of 6 Bohr  magnetons and, in consequence, the intensity of
the ensuing magnetic dipole--dipole interactions among the atoms is
much larger than the corresponding for alkali atoms. The interest in
dipolar degenerate quantum gases lies in the fact that novel
phenomena are expected to emerge, For instance, its expansion, after
releasing it from an anisotropic trap, entails that the anisotropy
of the dipole--dipole interaction can be tracked down to a
detectable anisotropic deformation of the expansion of the system
\cite{Stuhler1}. Clearly, additional effects shall emerge, as an
example we may add that this dipole--dipole interaction, which has a
long range, shall modify the shape of the condensate
\cite{Giovinazzi1}. It should be no surprise that this
dipole--dipole interaction must modify the speed of sound in this
kind of systems. Indeed, the absence of interatomic interactions
means that sound cannot exist \cite{Pitaevski1}. In other words,
interatomic interactions are a key ingredient in the definition of
the speed of sound, therefore, the presence of an additional force,
this magnetic dipole--dipole feature, should impinge upon the speed
of sound. The analysis of the speed of sound allows us to include
anisotropy as an additional characteristic in this physical
parameter.

In the present work we consider a BEC in which the trapping
potential is endowed with the mathematical form of an anisotropic
harmonic oscillator and, in addition, we take into account the
presence of a magnetic dipolar moment such that a dipole--dipole
interaction emerges as a key element in the description of our
system. Some parameters are deduced, for instance, speed of sound,
radii of the BEC, chemical potential, and number of particles. An
interesting point is related to the fact that the speed of sound is
anisotropic.

\section{Dipole--Dipole Interaction and Condensation}

\subsection{Dipole--Dipole Mean Field Theory Energy}

The system under study is a  gas of chromium atoms in which an
appropriate rotating magnetic field entails the emergence of dipolar
interactions among the atoms of the system. This kind of
interactions are long--ranged and imply the presence of anisotropy.
These two features are in contrast with the properties of those
pertaining to a short--range and isotropy interaction contained in
the so--called scattering length. We introduce, from the very
beginning, the Thomas--Fermi limit, and, within this context, we
deduce the chemical potential, radii of the condensate, number of
particles, and speed of sound.

Consider two equal atoms, one located at $\vec{r}$ and the second
one at $\vec{R}$, then a potential energy related to a
dipole--dipole interaction between them appears, here $\gamma$
denotes the gyromagnetic ratio \cite{Cohen2}

\begin{eqnarray}
V_d(\vec{r}-\vec{R})=\frac{\mu_0\gamma^2}{4\pi
\vert\vert\vec{r}-\vec{R}\vert\vert^3}\Bigl[\vec{S_1}\cdot\vec{S_2}\nonumber\\
-3\frac{\bigl(\vec{S_1}
\cdot(\vec{r}-\vec{R})\bigr)\bigl(\vec{S_1}\cdot(\vec{r}-\vec{R})\bigr)}{\vert\vert\vec{r}-\vec{R}\vert\vert^2}\Bigr].
\label{equa1}
\end{eqnarray}

An external magnetic field polarizes all our chromium atoms along
the $z$--axis, such that

\begin{equation}
\vec{S_1} = \vec{S_2} =3\hbar\vec{e_z}.\label{equa2}
\end{equation}

Under these circumstances the dipolar potential reads

\begin{equation}
V_d(\vec{r}-\vec{R})=\frac{\mu_0(6\mu_B)^2}{4\pi}
\frac{1}{\vert\vert\vec{r}-\vec{R}\vert\vert^3}\Bigl[1-3\frac{(z-Z)^2}{\vert\vert\vec{r}-\vec{R}\vert\vert^2}\Bigr].
\label{equa3}
\end{equation}

This energy will be considered in the context of the mean field
theory, namely,

\begin{equation}
E_d= \frac{1}{2}\int
d\vec{r}d\vec{R}V_d(\vec{r}-\vec{R})n(\vec{r})n(\vec{R}).\label{equa4}
\end{equation}

The time--independent Gross--Pitaevskii equation reads

\begin{eqnarray}
\mu\psi(\vec{r}) = -\frac{\hbar^2}{2m}\nabla^2\psi(\vec{r})+
V_t((\vec{r})\psi(\vec{r}) +\nonumber\\
U_0\vert\psi(\vec{r})\vert^2\psi(\vec{r})+ \int
d\vec{R}V_d(\vec{r}-\vec{R})n(\vec{R})\psi(\vec{r}). \label{equa5}
\end{eqnarray}

Here $U_0=4\pi a\hbar^2/m$ \cite{Ueda1} ($a$ is the scattering
length) and $V_t$ denotes the trapping potential and for our case we
have that it depicts an anisotropic harmonic oscillator

\begin{equation}
V_t((\vec{r}) = \frac{m}{2}\Bigl[\omega^2_0x^2 + \omega^2_0y^2 +
\omega^2_z\Bigr].\label{equa6}
\end{equation}

\subsection{Chemical Potential and Geometry of the Condensate}

At this point we introduce two simplifications, namely; (i) the
Thomas--Fermi approximation, i.e., we neglect the kinetic term; (ii)
in the integral term of (\ref{equa5}) we consider
$n(\vec{r}´)\Rightarrow n^{(0)}(\vec{r}) =(\mu-V_t(\vec{r}))/U_0$,
in other words, for the calculation of this integral term we
introduce for the density the corresponding function appearing in
the case of vanishing dipole--dipole interaction \cite{Pethick1}. In
order to simplify the calculations let us point out the following
identity

\begin{equation}
\frac{1}{\vert\vert\vec{r}-\vec{R}\vert\vert^3}-
3\frac{(z-Z)^2}{\vert\vert\vec{r}-\vec{R}\vert\vert^5}=
-\frac{\partial^2}{\partial
z^2}\Bigl[\frac{1}{\vert\vert\vec{r}-\vec{R}\vert\vert}\Bigr]
-\frac{4\pi}{3}\delta(\vec{r}-\vec{R}).\label{equa7}
\end{equation}

This last expression entails

\begin{equation}
\int d\vec{R}V_d(\vec{r}-\vec{R})n^{(0)}(\vec{R})
=-\mu_0(6\mu_B)^2\Bigl[\frac{1}{3}n^{(0)}(\vec{r})+
\frac{\partial^2}{\partial z^2}(\phi(\vec{r}))\Bigr], \label{equa8}
\end{equation}

where

\begin{equation}
\phi(\vec{r})=
\frac{1}{4\pi}\int\frac{n^{(0)}(\vec{R})}{\vert\vert\vec{r}-\vec{R}\vert\vert}d^3R.
\label{equa9}
\end{equation}

We must now provide the volume of integration. In the absence of
dipole--dipole interaction and with an isotropic trap, i.e.,
$\omega_0=\omega_z$, the volume is a sphere whose radius $R$ can be
obtained, within the context of the Thomas--Fermi approximation,
imposing the condition

\begin{equation}
n(r=R)=0\Rightarrow\mu=V_t(r=R). \label{equa10}
\end{equation}

The value of the chemical potential equals the value of the trap
evaluated at the boundary of the condensate. In the case of an
anisotropic trap we do not expect to have a sphere \cite{Pethick1}.
For a situation like the one considered here, in connection with the
trap, the geometry of the condensate should be derived, as in the
case of an isotropic trap, from our model. Clearly, we expect to
have, due to the symmetry of the trap, to equal sizes $R_0$ and a
second one different $R_z$. These parameters do not determine,
completely, the geometry of the system. The presence of a
dipole--dipole interaction along a certain direction complicates
even more the question of the corresponding geometry. Indeed, this
kind of interactions may change, drastically, the geometry of the
condensate \cite{Santos1}.

The deduction of some of the properties of the condensate requires
the knowledge of the geometry of the system, for instance, we must
calculate the integral appearing in (\ref{equa9}), a fact that needs
the integration volume. For our case we will assume that the system
has an ellipsoidal geometry, with two equal axes $R_0$, along the
$x$ and $y$ axes, and the third one different $R_z$.

Our coordinate system will be chosen to be prolate spheroidal
\cite{Arfken1}

\begin{eqnarray}
x= R_z\sinh u\cos v\cos\phi,\nonumber\\
y= R_z\sinh u\cos v\sin\phi,\nonumber\\
z= R_z\cosh u\sin v. \label{equa11}
\end{eqnarray}

In these last expressions we have that $0\leq v\leq \pi$; $0\leq
\phi\leq 2\pi$; $0\leq u\leq\tilde{u}$; $tanh(\tilde{u})=R_0/R_z$.

Consider now a point within the condensate with position vector
$\vec{r}$, then

\begin{eqnarray}
\frac{1}{4\pi}\int\frac{\mu}{\vert\vert\vec{r}-\vec{R}\vert\vert}d^3R
=
\frac{1}{4\pi}\int_{V_1}\frac{\mu}{\vert\vert\vec{r}-\vec{R}\vert\vert}d^3R\nonumber\\
+
\frac{1}{4\pi}\int_{V_2}\frac{\mu}{\vert\vert\vec{r}-\vec{R}\vert\vert}d^3R.
\label{equa12}
\end{eqnarray}

Here $V_1$ denotes a sphere of radius $r$ centered at the origin of
the condensate, whereas $V_2$ denotes the remaining volume of the
condensate.

Clearly, the first integral on the right--hand side is given by

\begin{equation}
\frac{1}{4\pi}\int_{V_1}\frac{\mu}{\vert\vert\vec{r}-\vec{R}\vert\vert}d^3R
= \frac{\mu}{3}r^2. \label{equa13}
\end{equation}

The second integral takes the following form

\begin{eqnarray}
\frac{1}{4\pi}\int_{V_2}\frac{\mu}{\vert\vert\vec{r}-\vec{R}\vert\vert}d^3R
=\nonumber\\
2\pi\mu
R^2_z\Bigl[\frac{2}{3}\Bigl(1\bigl(\frac{r}{R_z}\bigr)^2\Bigr)\Bigl(\cosh^3\tilde{u}-\cosh\tilde{u}-\nonumber\\\bigl(\frac{r}{R_z}\bigr)^3+\bigl(\frac{r}{R_z}\bigr)\Bigr)
-\frac{z\pi}{4R_z}\Bigl(\cosh^4\tilde{u}-\cosh^2\tilde{u}-\nonumber\\
\bigl(\frac{r}{R_z}\bigr)^4+\bigl(\frac{r}{R_z}\bigr)^2\Bigr)\Bigr].\label{equa15}
\end{eqnarray}

Similarly for the remaining integrals. After a messy calculation it
can be proved that the mean field version of the dipole--dipole
interaction is provided by

\begin{eqnarray}
\int
d\vec{R}V_d(\vec{r}-\vec{R})n^{(0)}(\vec{R})=-\frac{\mu_0(2\mu_bm_s)^2}{U_0}\Bigl[\frac{\mu}{3}\nonumber\\
-\frac{m}{6}\Bigl(\omega^2_0x^2+\omega^2_0y^2+ \omega^2_zz^2\Bigr)
+\mu\frac{R_z\cosh(\tilde{u})-r}{3R_z}\nonumber\\
-\frac{m\omega^2_0R^2_z}{4}\Bigl(\frac{2R_z\cosh(\tilde{u})-2r}{5R_z}+
\frac{zr\pi}{32R^2_z}(1+\frac{z^2}{r^2})\Bigr)\nonumber\\
-\frac{4m\omega^2_zR^2_z}{105}(1+3\frac{z^2}{r^2})(1-\frac{z^2}{2r^2})(\frac{r}{R_z})^3.\label{equa16}
\end{eqnarray}

For the sake of brevity, from now on we will use

\begin{eqnarray}
\gamma = \frac{\mu_0(2\mu_bm_s)^2}{U_0}.\label{equa198}
\end{eqnarray}

The chemical potential can be obtained from the fact that the
density shall vanish at any point on the boundary surface of the
condensate. Indeed,

\begin{eqnarray}
n(\vec{r}) = \frac{1}{U_0}\Bigl[\mu-V_t(\vec{r})-\int
V_d(\vec{r}-\vec{R})n(\vec{r})n(\vec{R})d\vec{R}\Bigr]
.~~~\label{equa178}
\end{eqnarray}

In other words, the condition $n(\vec{r}=R_0\vec{e}_x)=0$ implies

\begin{eqnarray}
\mu=\Bigl(1+\frac{\gamma}{3}(1+\cosh(\tilde{u})-\frac{R_0}{R_z})\Bigl)^{-1}\Bigl\{
\frac{m\omega^2_0R^2_0}{2}\nonumber\\
+\gamma\Bigl[\frac{m\omega^2_0R^2_0}{6} +\frac{m\omega^2_0R^2_z}{2}
\bigl(\frac{2}{5}[\cosh\tilde{u}-\frac{R_0}{R_z}]\bigr)\nonumber\\
+4\frac{m\omega^2_zR^2_z}{105}\frac{2R^2_zR^3_0-R^5_0}{2R^5_z}
\Bigr]\Bigr\} .\label{equa17}
\end{eqnarray}

We may find a relation between $R_0$ and $R_z$ recalling that
$n(\vec{r}=R_z\vec{e}_z)=0$ provides also the chemical potential.

\begin{eqnarray}
\mu=\Bigl(1+\frac{\gamma}{3}(\cosh(\tilde{u})-3\pi)\Bigl)^{-1}\Bigl\{
\frac{m\omega^2_zR^2_z}{2}\nonumber\\
+\gamma\Bigl[\frac{m\omega^2_zR^2_z}{6} +\frac{m\omega^2_0R^2_z}{2}
\bigl(\frac{2}{3}[\cosh\tilde{u}-1]\nonumber\\
+\frac{\pi}{16}\bigr)
+2\frac{m\omega^2_zR^2_z}{105} \Bigr]\Bigr\}.\label{equa18}
\end{eqnarray}

The comparison between these two last expressions renders a
transcendental equation which determines the ratio $R_0/R_z$ as a
function $\omega_0$, $\omega_z$, and $\gamma$.

\begin{eqnarray}
\Bigl(\frac{\omega_0R_0}{\omega_zR_z}\Bigr)^2=
\frac{1+\frac{\gamma}{3}[1+\cosh(R_0/R_z)-R_0/R_z]}
{1+\frac{\gamma}{3}[\cosh(R_0/R_z)-3\pi]}\times\nonumber\\
\Bigl\{1+\frac{\gamma}{3}[1+(\omega_0/\omega_z)^2[2\cosh(R_0/R_z)-2+\pi/2]\Bigr\}\times\nonumber\\
\Bigl\{1+\frac{\gamma}{3}[1+\frac{3R^2_z}{5R^2_0}[2\cosh(R_0/R_z)-2R_0/R_z\nonumber\\
+\frac{8R_0\omega^2_z}{7R_z\omega^2_0}(1-\frac{R^2_0}{2R^2_z})]\Bigr\}^{-1}.~~~\label{equa19}
\end{eqnarray}

This last expression will be satisfied only for certain values of
$R_0/R_z$ (assuming $\omega_0$, $\omega_z$, and $\gamma$ are known)
but it does not provide the value of $R_0$ (or of $R_z$). In order
to obtain these parameters we require an additional equation.

\subsection{Speed of Sound, Number of Particles, and energy per particle}

The speed of sound ($c_s$) is given by $c^2_s=
\frac{n}{m}(\frac{\partial\mu}{\partial n})$ \cite{Pathria1}. For
our particular case we have ($\delta=R_0/R_z$)

\begin{eqnarray}
c_s^2=\frac{\omega^2_zR^2_z}{2}\Bigl\{1+\frac{\gamma}{3}\Bigl[\cosh(\delta)-3\pi\Bigr]\Bigr\}^{-1}\nonumber\\
\Bigl\{1+\frac{\gamma}{3}[1+(\omega_0/\omega_z)^2[2\cosh(R_0/R_z)-2+\pi/2]\Bigr\}\nonumber\\
-\frac{\omega^2_0x^2+\omega^2_0y^2+\omega^2_zz^2}{2}+\frac{\gamma}{m}\Bigl\{\frac{\mu}{3}-\nonumber\\
\frac{m}{6}\Bigl(\omega^2_0x^2+\omega^2_0y^2+ \omega^2_zz^2\Bigr)
+\frac{\mu}{3}\Bigl[\cosh(\delta)-\frac{r}{R_z}-\frac{3\pi
z}{R_z}\Bigr]\nonumber\\
-\frac{m\omega^2_0R^2_z}{4}\Bigl[\frac{2}{5}\bigl(\cosh(\delta)-\frac{r}{R_z}\bigr)+\frac{r\pi
z}{32R^2_z}\bigl(1+\frac{z^2}{r^2}\bigr)\Bigr]\nonumber\\
-2\frac{m\omega^2_zR^2_z}{105}\bigl(1+\frac{3z^2}{r^2}\bigr)\bigl(\frac{2r^3R^2_z-r^5}{r^5_z}\bigr)\Bigr\}.~~~~.\label{equa20}
\end{eqnarray}

This last expression provides us with the possibility of deducing
the geometrical parameters of the condensate through the value of
the speed of sound at the center of the condensate, i.e.,
$c_s^2(\vec{r}=0)=c_s^2(0)$. Indeed,

\begin{eqnarray}
R^{-2}_z= \frac{\omega^2_zc_s^2(0)}{2}\Bigl\{1+\frac{\gamma}{3}\Bigl[\cosh(\delta)-3\pi\Bigr]\Bigr\}^{-1}\nonumber\\
\Bigl\{1+\frac{\gamma}{3}[1+(\omega_0/\omega_z)^2[2\cosh(\delta)-2+\pi/2]\Bigr\}\nonumber\\
\Bigl\{1+\frac{\gamma}{3}\Bigl[1+\cosh(\delta)\Bigr]\Bigr\}
\frac{\omega^2_0c_s^{-2}(0)}{2}\cosh(\delta).\label{equa21}
\end{eqnarray}

This expression provides the value of $R_z$ as a function of
measurable parameters.

The number of particles can be obtained integrating (\ref{equa178}),
and we have, approximately

\begin{eqnarray}
N=\frac{4\pi\mu}{3U_0}\Bigl\{1+\frac{\gamma}{3}\Bigr\}R^2_0R_z-\Bigl\{1+\frac{\gamma}{3}\Bigr\}\frac{2m\pi\omega^2_zR^2_z}{15U_0}\nonumber\\
\Bigl\{\Bigl[2+(\frac{\omega_0)}{\omega_z})^2\Bigr]\cosh^5(\delta)-\frac{2}{3}\Bigl[4(\frac{\omega_0}{\omega_z})^2-1\Bigr]\cosh^3(\delta)\nonumber\\
3(\frac{\omega_0}{\omega_z})^2\cosh(\delta)-\frac{4}{3}\Bigl[(\frac{\omega_0}{\omega_z})^2+1\Bigr]\Bigr\}.~~~.\label{equa22}
\end{eqnarray}

We know that $\mu= (\frac{\partial E}{\partial N})_{(T,V)}$
\cite{Pathria1}, therefore we may find the internal energy of the
condensate. Indeed, $E=\int\mu dN = \int\mu\frac{dN}{d\mu}d\mu$, and
therefore, after a lengthy calculation we find that

\begin{eqnarray}
E=\frac{3}{2}\Bigl\{\Bigl(\frac{a_1+a_2a_3}{a_2^2}\Bigr)\Bigl(\frac{2}{7}z^{7/2}\nonumber\\
-\frac{4}{5}bz^{5/2}+\frac{2}{3}b^2z^{3/2}\Bigr)
+\Bigl(\frac{a_3bz^{3/2}}{a_2}\Bigr)\Bigl(\frac{2}{5}z
-\frac{2}{3}b\Bigr)\nonumber\\
+\frac{a_1+a_2a_3}{a_2}\Bigl(\frac{2}{7}z^{7/2}-\frac{2}{5}bz^{5/2}\Bigr)\Bigr\}
.~~~~~~\label{equa23}
\end{eqnarray}

In these expressions we have the following parameters:

\begin{eqnarray}
z=a_2\mu + b,~~b=-mc^2_s(0),a_2=1+\frac{\gamma}{3}\Bigl[1+\cosh(\delta)\Bigr],\nonumber\\
a_1=\frac{4\pi\delta^2}{3U_0}\Bigl[1+\frac{\gamma}{3}\Bigr]\Bigl(\frac{10}{m\omega^2_0\cosh(\delta)}\Bigr)^{3/2},~~~~
,\label{equa24}
\end{eqnarray}

\begin{eqnarray}
a_3=\Bigl[1+\frac{\gamma}{3}\Bigr]\Bigl(\frac{10}{m\omega^2_0\cosh(\delta)}\Bigr)^{5/2}\Bigl\{\Bigl(2
+\bigl(\frac{\omega_0}{\omega_z}\bigr)^2\Bigr)\cosh^5(\delta)\nonumber\\
-\frac{2}{3}\Bigl(4\bigl(\frac{\omega_0}{\omega_z}\bigr)^2-1\Bigr)\cosh^3\delta)+3\bigl(\frac{\omega_0}{\omega_z}\bigr)^2\cosh(\delta)\nonumber\\
-\frac{4}{3}\Bigl(\bigl(\frac{\omega_0}{\omega_z}\bigr)^2-1\Bigr).~~~~~
.\label{equa224}
\end{eqnarray}

Clearly, this last result allows us to find the average energy, per
particle, $\epsilon=E/N$.

\section{Discussion}

We have considered a bosonic system comprised by chromium atoms in
which the corresponding trap has two equal frequencies. In addition,
a magnetic field polarizes all the chromium atoms and, in
consequence, dipole--dipole interactions emerge as an important
element in the physics of the system. Concerning the chemical
potential in this scheme we have that it is given by (\ref{equa17}).
It is readily seen that the chemical potential is a non--linear
function of the parameters of the trap and the scattering length, a
fact which is no surprise, since previous results have shown this
kind of dependence \cite{Pathria1}. At this point we must underline
that in these aforementioned examples it has been assumed that the
strength of the contact interaction is larger than that stemming
from the dipole--dipole term. In the present situation this
assumption has not been imposed and, in this sense, it provides a
more general case. Of course, we may recover the results of these
previous works. Indeed, consider, for instance, the case of
vanishing dipole--dipole interaction and an isotropic harmonic
oscillator. Under these conditions the Thomas--Fermi approximation
defines the following relation among the radius of the condensate
($R_i$), frequency ($\omega_i$), and chemical potential ($\mu$),
namely, $\mu=\frac{\omega_iR_i^2}{2}$, where $i=0, z$. Expressions
(\ref{equa17}) and (\ref{equa18}) imply this value of the chemical
potential if we set $\gamma=0$. Of course, in the present situation
the dependence of the chemical potential upon the frequencies,
scattering length, and dipole--dipole interaction is more
complicated. Since the value of the chemical potential in the region
of temperatures between $T=0$ and $T=T_c$  is a constant and it
coincides with the energy of the ground state then we know that
$\epsilon_0 = \mu$.

Clearly, (\ref{equa19}) cannot provide us with the value of $R_0$ or
$R_z$, but only with the ratio $R_0/R_z$. This fact is no surprise
at all and it is also present in the simplest case in which the
dipole--dipole interaction and the contact interaction are switched
off. Indeed, for this last case we know that there is no diffusion
of particles within the system when the chemical potential has the
same value at all points, hence
$\omega^2_0R^2_0=\omega^2_zR^2_z\Rightarrow\omega^2_0/\omega^2_z=R^2_z/R^2_0$.
In other words, thermodynamical arguments do not determine $R_0$ or
$R_z$, only its ratio. For our situation, given the frequencies and
$\gamma$, (\ref{equa19}) provides us the ratio $R_0/R_z$ which is
unique, namely, there is only one case in which $R_0$ and $R_z$ are
both positive and fulfill this foresaid expression. In order to
determine the value of $R_0$ or $R_z$ we may resort to expression
(\ref{equa21}) where we may deduce $R_z$ as a function of the speed
of sound at the center of the condensate, of $\delta$, frequencies,
etc.

Concerning the speed of sound in this system (\ref{equa20}), it is
readily seen that it is not only position--dependent but also
anisotropic. To fathom better this last statement notice that
(\ref{equa20}) implies that the motion along the $z$--axis happens
at a different speed that along the $x$--axis. Indeed, if in
(\ref{equa20}) we impose the condition $y=z=0$ and calculate the
corresponding speed of sound, and afterwards we do the same, but
with the condition $x=y=0$, the results differ. The reason for this
lies not only on the fact that the frequencies along $x$ and $z$ are
different (see the term $\omega_0^2x^2+\omega_0^2y^2+\omega_z^2z^2$
in (\ref{equa20})) but also on the fact that there is dipole--dipole
interaction only along the $z$-axis. For instance, the speed of
sound includes the term $-\frac{\gamma\mu}{3m}\frac{3\pi z}{R_z}$
but it does not include terms of the form
$-\frac{\gamma\mu}{3m}\frac{3\pi x}{R_0}$ or
$-\frac{\gamma\mu}{3m}\frac{3\pi y}{R_0}$. This absence can be
comprehended as a consequence of a dipole--dipole interaction only
along the $z$--axis.

The possible relevance of this anisotropic behavior of the speed of
sound is related to fact that it could provide us with a tool to
investigate some dissipative mechanisms, in particular the case of
the Landau criterion for superfluidity \cite{Nozieres1} seems to be
a feasible case \cite{Giovinazzi1}. Notice that our calculation
confirms the conjecture mentioned in \cite{Giovinazzi1} about the
loss of isotropy in the speed of sound for a chromium BEC, and, in
consequence, makes sounder the possibility of studying dissipative
mechanisms with systems in which dipole--dipole interactions play a
relevant role in the definition of its physical properties.

\end{document}